\documentclass[aps,pra,twocolumn,amsmath,showpacs]{revtex4}
\usepackage{graphicx}
\usepackage{amsmath}
\begin{document}
\title{Diatomic molecules in ultracold Fermi gases - Novel composite
bosons}
\author{D.S. Petrov$^1$,  C. Salomon$^2$, and G.V.
Shlyapnikov$^{3,4}$}
\affiliation{$^1$ITAMP, Harvard-Smithsonian Center for Astrophysics,
and Harvard-MIT
Center for Ultracold Atoms,
Department of Physics, Harvard University, Cambridge, Massachusetts
02138, USA\\
$^2$Laboratoire Kastler Brossel, 24 rue Lhomond, F-75231 Paris
CEDEX, France\\
$^3$Laboratoire Physique Th\'eorique et Mod\`eles Statistique,
Universit\'e Paris
Sud, B\^at. 100, 91405 Orsay
CEDEX, France\\
$^4$Van der Waals-Zeeman Institute, University of Amsterdam,
Valckenierstraat 65/67, 1018 XE
Amsterdam, The Netherlands}
\date{\today}
\begin{abstract}
We give a brief overview of recent studies of weakly bound
homonuclear molecules in ultracold two-component Fermi gases. It is
emphasized that they represent novel composite bosons, which
exhibit features of Fermi statistics at short intermolecular
distances. In particular, Pauli exclusion principle for identical
fermionic atoms provides a strong suppression of collisional
relaxation of such molecules into deep bound states. We then 
analyze heteronuclear molecules which are 
expected to be formed in mixtures of different fermionic atoms. 
It is found how an increase in the mass ratio for the constituent 
atoms changes the physics of collisional stability of such molecules 
compared to the case of homonuclear ones. We discuss
Bose-Einstein condensation of these composite bosons and draw
prospects for future studies.

\end{abstract}

%
\maketitle

\section{Introduction} \label{sec.int}

Quantum statistics for bosons and fermions  lies in the basis of
our understanding of many-body physics. In the case of a dilute
gas of bosons, one faces the phenomenon of Bose-Einstein
condensation (BEC), a macroscopic occupation of a single quantum
state below a certain critical temperature, predicted by S. Bose
and A. Einstein in 1924 \cite{B,E}.  The idea of BEC originally
comes from the analogy between light and matter waves. Indeed, at
sufficiently low temperatures the de Broglie wavelength of
particles exceeds the mean inter-particle separation, and the wave
packets of particles start overlapping. Then the quantum
statistics comes into play and it becomes favorable for bosons to
occupy a single quantum state. This state represents a macroscopic
quantum object called Bose-Einstein condensate, and the formation
of the condensate manifests itself as a phase transition
accompanied by a change of some of the properties of the
gas.

Since the work of Bose \cite{B} and Einstein \cite{E}, a number of
phenomena in interacting systems have been considered as
manifestations of BEC: superfluidity in liquid helium, high-$T_c$
superconductivity in some materials, condensation of hypothetical
Higgs particles, BEC of pions, and so on. Bose-Einstein
condensation in dilute gases has been  first observed in 1995 in
pioneering experiments with clouds of magnetically trapped alkali
atoms at JILA \cite{BEC1}, MIT \cite{BEC2}, and Rice \cite{BEC3}.
The first generation of studies of trapped Bose-condensed gases
found a variety of spectacular macroscopic quantum effects:
interference between two condensates, collective
oscillations and their damping, formation of quantized vortices
and vortex lattices, atom lasers, etc. (see \cite{Lev} for
review).

In the last few years, the field of quantum gases is strongly
expanding in the direction of ultracold two-component clouds of
fermionic atoms, with the initial goal of achieving a superfluid
Bardeen-Cooper-Schrieffer (BCS) phase transition. This transition
requires attractive interaction between the atoms. Then, in the
simplest version of the superfluid transition, at sufficiently low
temperatures fermions belonging to different components and having
opposite momenta on the Fermi surface form correlated (Cooper)
pairs in the momentum space. This leads to the appearance of a gap
in the excitation spectrum and to the phenomenon of superfluidity
(see, for example, \cite{LL}). In a dilute ultracold two-component
Fermi gas, most efficient is the formation of Cooper pairs due to
the attractive intercomponent interaction in the s-wave channel
(negative $s$-wave scattering length $a$). However, for ordinary
values of $a$, the superfluid transition temperature is extremely
low.

For this reason, the efforts of experimental groups have been
focused on modifying the intercomponent interaction by using
Feshbach resonances (see below). In this case, one can switch the
sign and tune the absolute value of $a$, which at resonance
changes from $+\infty$ to $-\infty$. This has led to remarkable
developments, in particular to the creation of weakly bound
diatomic molecules of fermionic atoms on the positive side of the
resonance ($a > 0$) \cite{ens1,rudy1,jila1}. These are the largest
diatomic molecules obtained so far. Their size is of the order of
$a$ and it reaches hundreds of nanometers in current experiments.
Accordingly, their binding energy is exceedingly small
($10^{-10}$\, eV or less). Being composite bosons, these molecules
obey Bose statistics, and they have been Bose-condensed in JILA
experiments with $^{40}$K$_2$ \cite{jila2,jila3} and in $^6$Li$_2$
experiments at Innsbruck \cite{rudy2,rudy3}, MIT \cite{mit1,mit2},
ENS \cite{ens2}, and Rice \cite{randy2}.

Nevertheless, some of the interaction properties of these
molecules reflect Fermi statistics of  the individual atoms
forming the molecule. In particular, these molecules are found
remarkably stable with respect to collisional decay. Being in the
highest rovibrational state, they do not undergo collisional
relaxation to deeply bound states on a time scale exceeding seconds
at densities of about $10^{13}$ cm$^{-3}$ , which is more than
four orders of magnitude longer than the life time of similar
molecules consisting of bosonic atoms.  

Currently, a new generation of experiments is being set up  
for studying degenerate mixtures of different fermionic atoms,
where one expects the formation of heteronuclear weakly bound 
molecules. In this paper we analyze elastic interaction between 
these molecules and their collisional relaxation to deeply bound
states.
The emphasis is put on the dependence of these properties on the
mass ratio for the constituent atoms and on the comparison
of the results with those for presently studied homonuclear molecules.
We present a physical picture of how such composite bosons can
exhibit features of Fermi statistics and show that this is the
reason for their collisional stability.

\section{Feshbach resonances and diatomic molecules in cold Fermi
gases}

The studies of interacting quantum gases are actively pursued
using Feshbach resonances, in particular for tuning and increasing
the intercomponent interaction in two-species Fermi gases. In the
vicinity of the resonance, for the two-particle problem one has a
strong coupling between (zero-energy) continuum states of
colliding atoms and a bound molecular state of another hyperfine
manifold. The effective coupling strength depends on the detuning
from the resonance (the energy difference $\delta$ between the
bound molecular state and the continuum states), which can be
varied by changing the magnetic field. Thus, the scattering length
becomes field dependent (see Fig.~1).
\begin{figure}
\includegraphics[width=8cm]{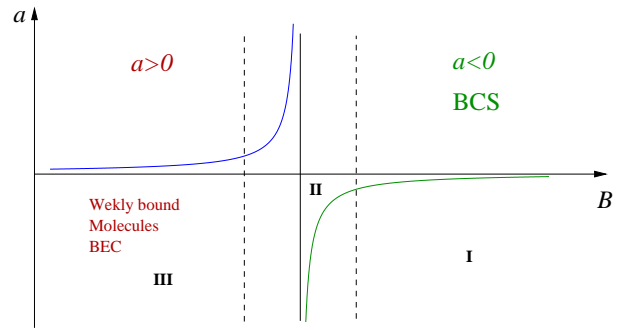}
\caption{The dependence of the scattering length on the magnetic
field. The symbols I, II, and III stand for the regime of a weakly
interacting atomic Fermi gas, strongly correlated regime, and the
regime of weakly bound molecules. At low enough temperatures region
I corresponds to the BCS superfluid pairing, and region III to
Bose-Einstein condensation of molecules.}
\end{figure}

At resonance the scattering length changes from $+\infty$ to
$-\infty$, and in the vicinity of the resonance one has the
inequality $k_F|a|\agt 1$ where $k_F$ is the Fermi momentum. The
gas then is in the strongly interacting regime. It is still dilute
and the mean interparticle separation greatly exceeds the
characteristic radius of interparticle interaction $R_e$. However,
the amplitude of binary interactions (scattering length) is larger
than the mean separation between particles, and the ordinary mean
field
approach is no longer valid.

For a large detuning from resonance the gas is still in the weakly
interacting regime, i.e. the inequality $n|a|^3\ll 1$ is
satisfied, where $n$ is the density of the gas. On the negative
side of the resonance ($a<0$), at sufficiently low temperatures of
a two-component atomic Fermi gas one expects the BCS pairing
between distinguishable fermions, well described in literature
\cite{LL}. On the positive side ($a>0$) two fermions belonging to
different components form weakly bound molecules. The collisional
stability and Bose-Einstein condensation of these molecules is a
subtle question, and it is the main topic of this paper.

The idea of resonance coupling through a Feshbach resonance for
achieving a superfluid phase transition in ultracold two-component
Fermi gases has been proposed in Refs. \cite{Holland1,Tim}. The
crossover from the BCS to BEC behavior attracts now a great deal
of interest, in particular with respect to the nature of
superfluid pairing, transition temperature, and elementary
excitations. It is worth noting that this type of crossover has
been discussed in literature in the context of superconductivity
\cite{Eag,Leg,Noz,Rand} and in relation to superfluidity in
two-dimensional films of $^3$He \cite{M,MYu}.

The description of a many-body system near a Feshbach resonance
requires a detailed knowledge of the 2-body problem. This is a
two-channel problem which can be described in terms of the
Breit-Wigner scattering \cite{Breit,LL3}, the open channel being
the states of colliding atoms and the closed channel the bound
molecular state of the other hyperfine domain. Various aspects of
this type of problems have been discussed by Feshbach
\cite{Feshbach} and Fano \cite{Fano}. In cold atom physics the
idea of Feshbach resonances was introduced in Ref. \cite{Verhaar},
and optically induced resonances have been discussed in Refs.
\cite{Gora,Bohn}.

The two-body physics is the most transparent if one can omit the
(small) background scattering length. Then for low collision
energies $\varepsilon$ the scattering amplitude is given by
\cite{LL3} :
\begin{equation} \label{FE}
F(\varepsilon)=-\frac{\hbar\gamma/\sqrt{2\mu}}
{\varepsilon+\delta+i\gamma\sqrt{\varepsilon}},
\end{equation}
where the quantity $\hbar\gamma/\sqrt{2\mu}\equiv W$ characterizes
the coupling between the two hyperfine domains and $\mu$ is the
reduced mass of the
two atoms. The scattering length is $a=-F(0)$. In Eq.~(\ref{FE}) the
detuning
$\delta$ is positive if the bound molecular state is below the
continuum of
colliding atoms. Then for $\delta>0$ the scattering
length is positive, and for $\delta<0$ it is negative. Introducing
a characteristic length
\begin{equation} \label{RW}
R^*=\hbar^2/2\mu W
\end{equation}
and expressing the scattering amplitude through  the relative
momentum of particles $k=\sqrt{2\mu\varepsilon}/\hbar$, Eq.~(\ref{FE})
takes the form:
\begin{equation} \label{Fk}
F(k)=-\frac{1}{a^{-1}+R^*k^2+ik},
\end{equation}
The validity of Eq.~(\ref{Fk}) does not require the inequality
$kR^*\ll 1$. At the same time, this equation formally coincides
with the amplitude of scattering of slow particles by a potential
with the same scattering length $a$ and an effective range
$R=-2R^*$, obtained under the condition $k|R|\ll 1$.

The length $R^*$ is an intrinsic parameter of the Feshbach
resonance problem. It characterizes the width of the resonance.
From Eqs.~(\ref{FE}) and (\ref{RW}) we see that large $W$ and,
hence, small $R^*$ correspond to a wide resonance, whereas small
$W$ and large $R^*$ lead to a narrow resonance. The issue of wide
and narrow resonances is now actively discussed in
literature
\cite{Combescot,Falco,Bruun,Bruun2,Petrovbosons,Palo,Eric,Ho3}.

In fact, the use of the terms ``wide'' and ``narrow'' depends on the
problem under consideration. For example, in the unitarity limit
where $a\rightarrow\pm\infty$, Eq.~(\ref{Fk}) shows  that the
length $R^*$ drops out of the problem under the condition $kR^*\ll
1$. In a quantum degenerate Fermi gas the characteristic momentum
of particles is the Fermi momentum $k_F=(3\pi^2n)^{1/3}$. Thus,
for a given $R^*$ the condition of the wide resonance depends on
the gas density $n$ and takes the form $k_FR^*\ll 1$
\cite{Bruun,Bruun2,Palo,Eric,Ho3}. Then the scattering properties of
the gas are the
same as in the case of potential scattering at an infinite scattering
length.

For $a>0$ and $na^3\ll 1$, where one has a weakly interacting gas
of diatomic bosonic molecules \cite{Petrov1}, the criterion of the
wide resonance is different \cite{Petrovbosons,PSS}. The most
important limitation is related to the binding energy and the
wavefunction of the molecules. The energy of the weakly bound
molecular state (it is certainly assumed that the characteristic
radius of interaction $R_e\ll a$) is determined by the pole  of
the scattering amplitude (\ref{Fk}). One then finds
\cite{Petrovbosons,PSS} that this state exists only for $a>0$ and
under the condition
\begin{equation} \label{Ra}
R^*\ll a.
\end{equation}
the binding energy is given by
\begin{equation}  \label{binding}
\varepsilon_0=\hbar^2/2\mu a^2.
\end{equation}
Then the wavefunction of the weakly bound molecular state has only
a small admixture of the closed channel, the size of the molecule
is $\sim a$, and atom-molecule and molecule-molecule interactions
are determined by a single parameter -- the atom-atom scattering
length $a$. In this sense, the problem becomes universal. It is
equivalent to the interaction problem for the two-body potential
which is characterized by a large positive scattering length $a$
and has a potential well with a weakly bound molecular state. The
picture remains the same when the background scattering length
$a_{bg}$ can not be neglected, although the condition of a wide
resonance can be somewhat modified \cite{Karen}.

Experimental studies are currently dealing with Fermi gases consisting
of atoms in two different internal (hyperfine) states and,
accordingly,
having equal masses. Most experiments use
wide Feshbach resonances. For example, weakly bound molecules
$^6$Li$_2$ and $^{40}$K$_2$ have been produced in experiments
\cite{ens1,rudy1,jila1,jila2,jila3,mit1,mit2,rudy2,rudy3,ens2,randy2}
by using Feshbach resonances with a length $R^*$ of the order of
or smaller than $20$\AA. Then, for the achieved values of the
scattering length $a$ from $500$ to $2000$\AA, the ratio $R^*/a$
is smaller than $0.1$. The only exception is the experiment at
Rice with $^6$Li near a narrow Feshbach resonance at 543 G
\cite{randy1}. For this resonance the length $R^*$ is very large
and at obtained values of $a$ the condition (\ref{Ra}) is not
fulfilled. Below we will focus attention on the case of a wide
Feshbach resonance.

\section{Elastic interaction between weakly bound
molecules\label{sec.elastic}}

At temperatures sufficiently lower than the molecular binding energy 
$\varepsilon_0$ and for equal concentrations of the two 
atomic components, practically all atoms are converted into
molecules if the gas density satisfies the inequality $na^3\ll 1$
\cite{Servaas}. This will be definitely the case at temperatures well
below the
Fermi energy (the least of the two Fermi energies in the case of atoms
with
different masses). Thus, one has a molecular Bose gas and the first
key question is the elastic interaction (scattering) between the
molecules. It is important for evaporative cooling of the
molecular gas to the regime of Bose-Einstein condensation and for
the stability of the condensate. In the latter respect, the
elastic interaction should be repulsive, otherwise the
Bose-condensed molecular gas undergoes a collapse.

The exact solution for the molecule-molecule elastic scattering
in the case of equal atom masses was found in Ref.~\cite{Petrov1}, 
and a detailed analysis of this
problem was given in Ref.~\cite{PSS}. This was done assuming
that the (positive) atom-atom scattering length $a$ for the
interspecies interaction greatly exceeds the characteristic radius
of interatomic potential:
\begin{equation}   \label{aRe}
a\gg R_e.
\end{equation}
Then, as in the case of the 3-body problem with fermions
\cite{STM,Danilov,efimov,Petrov2}, the amplitude of elastic
interaction is
determined only by $a$ and can be found in the zero-range
approximation for the interatomic potential.

This approach was introduced in the two-body physics by Bethe and
Peierls \cite{Bethe}. The leading idea is to solve the
equation for the free relative motion of two particles placing a
boundary condition on the wavefunction $\psi$ at a vanishing
interparticle distance $r$:
\begin{equation}\label{twobody.Bethe-Reierls}
\frac{(r\psi)'}{r\psi}=-\frac{1}{a},\,\,\,\,\,\,\,r\rightarrow 0,
\end{equation}
which can also be rewritten as
\begin{equation}\label{boundary1}
\psi\propto (1/r-1/a),\,\,\,\,\,\,\,r\rightarrow 0.
\end{equation}
One then gets a correct expression for the wavefunction at
distances $r\gg R_e$. For the case where $a\gg R_e$,
Eq.~(\ref{boundary1}) gives a correct result for the
wavefunction of weakly bound and continuum states even at
distances much smaller than $a$.

In this section we generalize the results on the molecule-molecule
scattering
obtained in
Refs.~\cite{Petrov1,PSS} to the case of molecules consisting of
fermionic atoms with
different masses $M$ and $m$ ($m<M$). This is relevant for weakly
bound heteronuclear
molecules, such as $^6$Li-$^{40}$K, $^6$Li-$^{87}$Sr,
$^{40}$K-$^{173}$Yb etc., 
which can be formed by sweeping across a
Feshbach resonance for the interspecies interaction. Elastic
interaction between
such molecules is important for understanding the physics of their
Bose-Einstein
condensation and for studying the BCS-BEC cross-over in Fermi
mixtures. 

The ultracold limit for the molecule-molecule scattering is realized
under the
condition
$ka\ll 1$, where  $k$ is the relative momentum. This is because $a$ is
approximately 
the range of interaction between the molecules. In this case the
scattering is
dominated by the contribution of the $s$-wave channel. The inequality
$ka\ll 1$ is
equivalent
to the collision energy much smaller than the molecular binding energy
$\varepsilon_0$.
Hence, the $s$-wave molecule-molecule elastic scattering can be
analyzed putting 
the total energy equal  to $-2\varepsilon_0=-\hbar^2/\mu a^2$.
In the zero-range approximation one should solve the four-body
free-particle
Schr\"odinger
equation:
\begin{eqnarray}\label{Poisson}
\left[-\nabla_{{\bf r}_1}^2-\nabla_{{\bf
r}_2}^2-\nabla_{\bf {R}}^2+2/a^2\right]\Psi =0,
\end{eqnarray}
placing the Bethe-Peierls boundary condition at vanishing
distances between heavy and light fermions. In Eq.~(\ref{Poisson}) 
the distance between two given heavy and light fermions is ${\bf
r}_1$, and ${\bf
r}_2$
is the distance between the other two (see Fig.~2). The distance
between the centers
of mass of
these pairs is $\beta {\bf R}$, and ${\bf r}_\pm=\alpha_\pm {\bf
r}_1+\alpha_\mp
{\bf r}_2\pm\beta {\bf R}$ are the separations between heavy and light
fermions in
the other two possible heavy-light pairs. Here we
introduced constants  
$\beta = \sqrt{2\alpha_+ \alpha_-}$, $\alpha_+=\mu/M$, $\alpha_-
=\mu/m$, and 
the reduced mass is $\mu=mM/(m+M)$. 

\begin{figure}
\label{fig.coordinates}
\includegraphics[width=0.7\hsize]{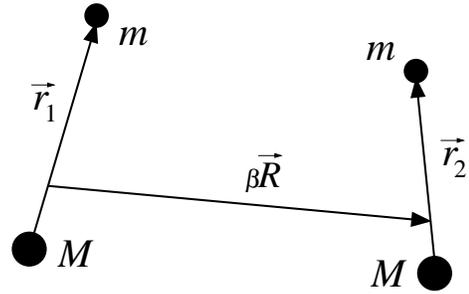}
\caption{Set of coordinates for the four-body problem.}
\end{figure}

The wavefunction $\Psi$ is symmetric with respect to the permutation
of composite
bosons and antisymmetric with respect to permutations of identical
fermions:
\begin{eqnarray}\label{symmetry}
&&\Psi({\bf r}_{1},{\bf r}_{2},{\bf {R}})=\Psi({\bf r}_{2},{\bf
r}_{1},-{\bf {R}})
\nonumber  \\
&&=-\Psi ({\bf r}_\pm,{\bf r}_\mp,\pm\beta ({\bf r}_1-{\bf
r}_2)\mp(\alpha_+
-\alpha_-){\bf R}).
\end{eqnarray}
The Bethe-Peierls boundary condition should be placed for a vanishing
distance 
in any pair of heavy and light fermions, i.e. for
${\bf r}_1\rightarrow 0$, ${\bf r}_2\rightarrow 0$, and ${\bf
r}_\pm\rightarrow 0$.
Due to the symmetry it is necessary to require a proper behavior of
$\Psi$ only at
one of these boundaries. For ${\bf r}_1\rightarrow 0$ the boundary
condition reads:
\begin{equation}\label{boundary}
\Psi({\bf r}_1,{\bf r}_2,{\bf {R}})\rightarrow f({\bf r}_2,{\bf
R})(1/4\pi r_1\,-1/4\pi a).
\end{equation}
The function $f({\bf r}_2,{\bf R})$ contains the information about the
second pair of
particles when the first two are on top of each other.

For large $R$ the wavefunction $\Psi$ is given by
\begin{equation}   \label{asymptote}
\Psi\approx\phi_0(r_1)\phi_0(r_2)(1-a_{dd}/\beta R);\,\,\,\,\,R\gg a,
\end{equation}
where $a_{dd}$ is the molecule-molecule scattering length, and the
wavefunction of
the
weakly bound molecule is given by 
\begin{equation}   \label{twobody.boundstate}
\phi_0(r)=(r\sqrt{2\pi a})^{-1}\exp (-r/a).
\end{equation}
Combining Eqs.~(\ref{boundary}) and (\ref{asymptote}) we obtain the
asymptotic
expression for $f$ at large distances $R$:
\begin{equation}\label{dimerdimer.swave}
f({\bf r}_2,{\bf R})\approx (2/r_2a)\exp{(-r_2/a)}(1-a_{dd}/\beta
R);\,\,\,\,R\gg a.
\end{equation}

For the $s$-wave scattering the function $f$ depends only on
three variables: the absolute values of ${\bf r}_2$ and ${\bf {R}}$,
and the angle
between them. An equation for the function $f$ in the case $m=M$ is
derived in
Refs.~\cite{Petrov1,PSS}. The generalization of this approach to the
case of different masses is
straightforward and we 
obtain the same integral equation:
\begin{eqnarray}\label{main}
&&\!\!\int\limits_{{\bf r}',{\bf {R}}'}\!\!\Big\{G(|{\bar
S}_1-S_1|)[f({\bf
r}',{\bf {R}}')
-f({\bf r},{\bf {R}})]+\bigl[G(|{\bar S}_1-S_2|)   \nonumber \\
&&\!\!-\sum_\pm G(|{\bar
S}_1-S_\pm|)\bigr]f({\bf r}',{\bf
{R}}')\Big\}=\frac{(\sqrt{2}-1)f({\bf r},{\bf {R}})}{4\pi a}.
\end{eqnarray}
The 9-dimensional vectors $S_1$, $S_2$, and ${\bar S}_1$ are again 
given by $S_1 = \{0,{\bf r}',{\bf {R}}'\},\;S_2 = \{{\bf
r}',0, -{\bf {R}}'\}$, ${\bar S}_1=\{0,{\bf r},{\bf {R}}\}$, 
and $G(x)=(2\pi)^{-9/2}(xa/\sqrt{2})^{-7/2}K_{7/2}(\sqrt{2}\,x/a)$ is
the Green function of Eq.~(\ref{Poisson}),  with $K_{7/2}$ being
the decaying Bessel function.  
The effect of different masses is contained in the expressions for the
vectors
$S_{\pm}$, which now read $S_\pm  =  \{\alpha_\mp {\bf r}' \pm \beta
{\bf
{R}}' ,\alpha_\pm {\bf r}' \mp\beta {\bf {R}}' , \mp\beta {\bf
r}' \mp(\alpha_+ -\alpha_-){\bf {R}}' \}$.

It is more convenient to make calculations in the momentum space,
transforming
Eq.~(\ref{main}) into an equation for the function $f({\bf k},{\bf
p})=\int d^3
rd^3Rf({\bf
r},{\bf R})\exp(i{\bf k\cdot r}/a+i\beta {\bf p\cdot R}/a)$:
\begin{eqnarray}\label{momentum}
\!\!\!\!\!&\!\!&\sum_\pm\!\!\int\!\!
\frac{f({\bf k}\pm \alpha_\mp ({\bf p}'-{\bf p}),{\bf p}')\,{\rm
d}^3p'}{2+\!\beta^2p'^2+\!({\bf k}\pm \alpha_\mp ({\bf p}'\!-{\bf
p}))^2+({\bf k}\pm
\alpha_\pm ({\bf p}'\!+{\bf p}))^2}\nonumber\\
\!\!\!\!&\!\!&\!=\!\int\!\! \frac{f({\bf k}',-{\bf p})\,{\rm
d}^3k'}{2\! +\! k'^2\!
+\! k^2\! +\! \beta^2p^2}- \frac{2\pi^2(1 +\! k^2\! +\beta^2p^2)f({\bf
k},{\bf
p})}{\sqrt{2 + k^2 +\beta^2p^2}+ 1}.
\end{eqnarray} 
By making the substitution $f({\bf k},{\bf p})=(\delta({\bf p})+g({\bf
k},{\bf
p})/p^2)/(1+k^2)$ we reduce Eq.~(\ref{momentum}) to an inhomogeneous
equation for
the function $g({\bf k},{\bf p})$. For ${\bf p}\rightarrow 0$ this
function tends to
a finite value
independent of ${\bf k}$. The molecule-molecule scattering length is
given by
$a_{dd}=-2\pi^2a\lim_{{\bf p}\rightarrow 0}g({\bf k},{\bf p})$, and we
have
calculated this
quantity numerically from Eq.~(\ref{momentum}). 

In Fig.~3 we display the ratio $a_{dd}/a$ versus the mass ratio $M/m$. 
For $m=M$ we recover the result of Refs. \cite{Petrov1,PSS}:
$a_{dd}=0.6a$. 
Compared to earlier studies which were
assuming $a_{dd}=2a$ \cite{Rand}, this exact result gives almost
twice as small a sound velocity of the molecular condensate and a
rate of elastic collisions smaller by an order of magnitude.
An approximate diagrammatic approach has been developed in Ref.
\cite{strinati},
and it has led to $a_{dd}=0.75a$.

The universal dependence of $a_{dd}/a$ on the mass ratio, presented in
Fig.~3,
can be established in the zero-range
approximation only if $M/m$ is smaller than approximately 13.6. 
Calculations then show the absence of four-body weakly bound states,
and
for $M/m\sim 1$ the behavior of $f$ suggests a soft-core repulsion
between
molecules, with a range $\sim a$.
For the mass ratio larger than the critical value, 
the description of the molecule-molecule scattering requires a
three-body
parameter
coming from the short-range behavior of the three-body subsystem
consisting of one
light
and two heavy fermions \cite{efimov,Petrov2}.     

\begin{figure}
\label{fig.add}
\includegraphics[width=\hsize]{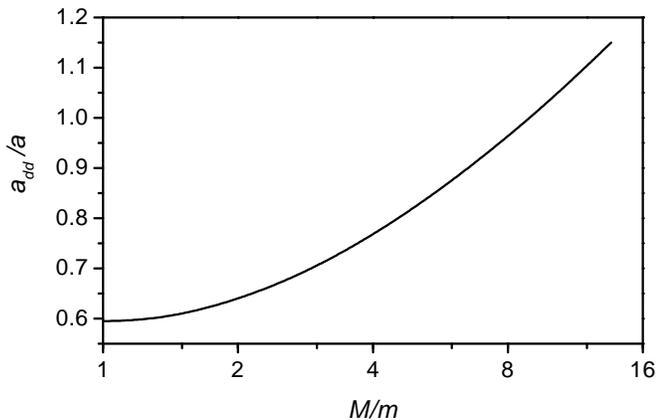}
\caption{The ratio $a_{dd}/a$ versus $M/m$.}
\end{figure}

\section{Born-Oppenheimer approach\label{sec.bo}}

In the case of large $M/m$ one can develop a transparent physical
picture of the molecule-molecule interaction by using the
Born-Oppenheimer approximation \cite{Fonseca}. In this case heavy
fermionic atoms are moving slowly in the field produced by the
exchange of fast light
atoms. The adiabatic behavior assumes that the four-body problem can
be split into
two parts. First, we calculate the wavefunctions and binding energies
of two light
fermions in
the field of two heavy atoms fixed at their positions ${\bf x}_1$ and
${\bf x}_2$. 
As the motion of heavy atoms is slow, the bound states of the light
atoms 
adiabatically adjust themselves to a given value of the separation
between 
the heavy ones, $x=|{\bf x}_1-{\bf x}_2|$. The sum of the
corresponding binding energies gives the effective interaction
potential
$U_{eff}(x)$ for the heavy fermions. An obvious second step is to
solve the
Shr\"odinger equation for their motion in this effective potential.

For $x>a$, there are two bound states of a light atom interacting with
a pair of
fixed heavy
atoms.
The wavefunctions of these states are given by
\begin{equation}\label{bo.boundstatelight}
\psi_{{\bf x},\pm}({\bf y})\propto\left(\frac{e^{-\kappa_\pm(x) |{\bf
y}-{\bf
x}_1|/x}}{|{\bf y}-{\bf x}_1|}\pm\frac{e^{-\kappa_\pm(x) |{\bf y}-{\bf
x}_2|/x}}{|{\bf y}-{\bf x}_2|}\right),
\end{equation}
where ${\bf y}$ is the coordinate of the light atom. 
The functions $\kappa_\pm (x)>0$ are determined by the requirement
that 
for ${\bf y}\rightarrow{\bf x}_{1,2}$ the wavefunction
(\ref{bo.boundstatelight})
satisfies the Bethe-Peierls boundary condition (\ref{boundary1}) in
which $r$ is
substituted by $|{\bf y}-{\bf
x}_{1,2}|$. This gives algebraic equations:
\begin{equation}\label{bo.lambda}
\kappa_\pm(x) \mp e^{-\kappa_\pm(x)}=x/a.
\end{equation}
The energies of the bound states are given by
\begin{equation}  \label{epsilon}
\varepsilon_\pm(x)=-\hbar^2\kappa_\pm^2(x)/2mx^2.
\end{equation}

Since the light fermions are identical, their two-body wavefunction
can be
constructed as an antisymmetrized product of $\psi_{{\bf x},+}$ and
$\psi_{{\bf
x},-}$:
\begin{equation}\label{bo.wavefunctionlight}
\psi_{\bf x}({\bf y}_1,{\bf y}_2)\propto\psi_{{\bf x},+}({\bf
y}_1)\psi_{{\bf
x},-}({\bf y}_2)-\psi_{{\bf x},+}({\bf y}_2)\psi_{{\bf x},-}({\bf
y}_1),
\end{equation} 
and their total energy is
$U_{eff}(x)=\varepsilon_+(x)+\varepsilon_-(x)$. Note that
the wavefunction (\ref{bo.wavefunctionlight}) is antisymmetric with
respect to the
transformation ${\bf x}\rightarrow -{\bf x}$. This means that solving
the
Shr\"odinger equation for the heavy fermions,
\begin{equation}\label{bo.schrodinger}
\left[-\frac{\hbar^2}{M}\nabla^2_{\bf x}+U_{eff}(x)\right]
\chi({\bf x})=E\chi({\bf x}),
\end{equation}
we have to look for a solution that is symmetric with respect to their
permutation
(in our case it corresponds to the $s$-wave scattering). Then the
total four-body
wave function $\Psi({\bf x}_1,{\bf x}_2,{\bf y}_1,{\bf y}_2)\propto
\chi({\bf
x})\psi_{\bf x}({\bf y}_1,{\bf y}_2)$ has a proper symmetry.

Solving Eqs.~(\ref{bo.lambda}) at distances $x>a$ gives a purely
repulsive potential
monotonically decreasing with $x$. At distances $x\gg a$ it behaves as
\begin{equation}   \label{Ueff}
U_{eff}(x)\approx
-2\varepsilon_0+(\hbar^2/2m)\exp(-2x/a)/ax;\,\,\,x\gg a. 
\end{equation}
We see that the
ratio of the $x$-dependent part of the effective repulsion to the
kinetic energy term
in
Eq.~(\ref{bo.schrodinger}) increases with $M/m$. This is consistent
with exact
calculations based on Eqs.~(\ref{main}) and (\ref{momentum}). 

The Born-Oppenheimer approach fails at distances close to $a$ as the
energy
$\varepsilon_-(x)$ vanishes at $x=a$, and the atom in the state
``$-$'' moves slower
than the heavy atoms. This  leads to a contradiction with the
adiabatic
approximation.
However, we see that the light atom in the state ``$-$'' becomes
essentially
delocalized and at distances $x\ll a$ the problem reduces to a
three-body problem
with two heavy fermions and one light fermion in the state ``$+$''. 
As has been mentioned, this problem is
characterized by a critical mass ratio $\approx 13.6$ above which the
behavior of
the system drastically changes. For our four-body problem this means
that the
effective potential at these distances becomes attractive and can
support bound
states. For an overcritical mass ratio, we also expect a resonance
dependence 
of molecule-molecule collisions on the scattering length or
short-range parameters 
of the system.  

Finally, it is worth noting that for a very large $M/m$ these
resonances should be
very narrow, since the repulsive effective potential at $x>a$ is very
strong.
Therefore,
the molecule-molecule scattering length should (on average) increase
logarithmically
with $M/m$.
Obviously, the effective repulsive barrier is important for the
analysis of the
inelastic losses in the molecular gas. We discuss this question in
more detail in
Sec.~VI.

\section{Suppressed collisional relaxation}

The most exciting physics with weakly bound bosonic molecules of
fermionic atoms is related to their collisional stability.
Actually, these are molecules in the highest rovibrational state
and they undergo relaxation into deeply bound states in
molecule-molecule (or molecule-atom) collisions. The released
binding energy of a deep state is $\sim\hbar^2/2\mu R_e^2$. It is
transformed into the kinetic energy of particles in the outgoing
collisional channel and they escape from the sample.  Thus,
collisional relaxation determines the lifetime of the Bose gas of
weakly bound molecules and is therefore a crucial process. Several
experiments show that such molecules consisting of bosonic
$^{87}$Rb \cite{heinzen,rempe} and $^{133}$Cs atoms \cite{rudy}
undergo a rapid collisional decay. On the other hand, first
observations of weakly bound molecules Li$_2$ and K$_2$,
consisting of fermionic atoms \cite{ens1,rudy1,randy1,jila2},
showed that they are long-lived at densities $\sim 10^{13}$
cm$^{-3}$.

We now arrive at the point where quantum statistics of composite
bosons comes into play for weakly bound molecules of fermionic
atoms. First, we discuss the case of $M/m\sim 1$, where the
interaction between molecules is characterized by a soft-core 
repulsion which does not significantly influence their behavior 
at short distances. Clearly behaving themselves as point-like bosons
at large
intermolecular distances, these molecules ``start remembering'' that
they
consist of fermions when the intermolecular separation becomes 
smaller than the molecule size ($\sim a$). As was explained in
Refs. \cite{Petrov1,PSS}, the 
key reason for the remarkable collisional stability of such weakly
bound molecules of fermionic atoms is Fermi statistics for the
atoms in combination with a large size of the molecular state
(small momenta of bound fermionic atoms). The physical picture is
the following. The binding energy of the molecules is
$\varepsilon_0=\hbar^2/2\mu a^2$ and their size is close to $a$. The
size of deep bound states is of the order of the characteristic
radius of interaction $R_e\ll a$. Hence, the relaxation requires
the presence of at least three fermions at distances $\sim R_e$
from each other. As two of them are necessarily identical, due to
the Pauli exclusion principle the relaxation probability acquires
a small factor proportional to a power of $(qR_e)$, where $q\sim
1/a$ is a characteristic momentum of the atoms in the weakly bound
molecular state.

The inequality $a\gg R_e$ allows one to obtain the dependence of
the molecule-molecule wavefunctions at short
interparticle distances on the two-body scattering length $a$ and
thus to establish a dependence of the relaxation rate on $a$. 
In Refs.~\cite{Petrov1,PSS} we have done this for $m=M$ and  
found a strong decrease of relaxation with increasing $a$.
Here we generalize these findings to the case of $m\neq M$,
assuming that the mass ratio $M/m$ is smaller than the critical
value $13.6$ and short-range physics does not influence the
$a$-dependence of the relaxation rate.

The relaxation is essentially a three-body process which occurs when
three fermions
approach each other to a short distance of the order of $R_e$. 
The fourth particle does not participate in the sense that the 
configuration space contributing to the relaxation
probability can be viewed as a system of three atoms at short
distances $\sim R_e$
from each other and a fourth atom separated from this system by a
large distance
$\sim a$. In this case and also for any hyperradius of the
three-fermion system
$\rho\ll a$, the four-body wavefunction decomposes into a product: 
\begin{equation}\label{decomp}
\Psi=\eta({\bf z})\Psi^{(3)}(\rho,\Omega),
\end{equation}
where  $\Psi^{(3)}$ is the wavefunction of the three-fermion system, 
$\Omega$ is the set of hyperangles for these fermions,  ${\bf
z}$ is the distance between their center of mass and the fourth atom,
and the
function $\eta({\bf z})$ describes the motion of this atom. This
function is
normalized to unity and averaging over the motion of the fourth atom
we see that the
probability of relaxation is determined solely by the short-range
behavior of
the function $\Psi^{(3)}(\rho,\Omega)$.

In the case of fermionic atoms with different masses one has two
possible choices
of a three-body subsystem out of four fermions. The most important is
the relaxation 
in the system of one atom with the mass $m$ and two heavier atoms with
masses $M$.  

The function $\Psi^{(3)}$ for $\rho\ll a$ follows from the
three-body Schr\"odinger
equation with zero total energy and, hence, can be written as
$\Psi^{(3)}=A(a)\psi(\rho,\Omega)$,
where the function $\psi$ is independent of $a$. Assuming that the
inelastic
amplitude of relaxation is much smaller than the amplitude of elastic
scattering,
the dependence of the relaxation rate constant on $a$ is related only
to the
$a$-dependence of the function $\Psi^{(3)}$ and, hence, is given by 
\begin{equation}  \label{alphaA}
\alpha_{rel}\propto |A(a)|^2.
\end{equation}
For finding $\alpha_{rel}(a)$ it is sufficient to consider distances
where
$R_e\ll\rho\ll a$. At these distances we have
$\psi=\Phi(\Omega)\rho^{\nu-1}$
\cite{PSS}, 
and the function $\Psi^{(3)}$ becomes 
\begin{equation}\label{psi3}
\Psi^{(3)}=A(a)\Phi(\Omega)\rho^{\nu-1}.
\end{equation}
The prefactor
$A(a)$ can be found from the solution of the four-body problem in the
zero-range
approximation.

Quite elegantly, the $a$-dependence of $A(a)$ can be determined from
the following
scaling considerations. The scattering length $a$ is the only length
scale in our
problem and we can measure all distances in units of $a$. For example,
rescaling
$\rho=a\rho'$ and ${\bf z}=a{\bf z}'$, taking into account that
$\eta({\bf z})$ is
normalized to unity, and using $\Psi^{(3)}$ from Eq.~(\ref{psi3}) we
obtain that the
$a$-dependent coefficient in Eq.~(\ref{decomp}) equals
$A(a)a^{\nu-5/2}$. By
applying the same rescaling to Eq.~(\ref{asymptote}) with the account
of
Eq.~(\ref{twobody.boundstate}) we see that the same coefficient should
be
proportional to $a^{-3}$. Therefore, $A(a)\propto a^{-\nu-1/2}$ and
$\alpha_{rel}\propto a^{-s}$, where $s=2\nu+1$.

The exponent $s$ depends on the mass ratio and on the symmetry of the
three-body wave
function $\Psi^{(3)}$. The leading relaxation channel at large $a$
corresponds to the
smallest $\nu$. The analysis of the three-body wavefunctions for the
$s$-wave and
$p$-wave 
atom-molecule collisions has been performed in
Ref.~\cite{Petrov2}. The $p$-wave symmetry in the system of one light
and two heavy
fermions provides the smallest value of $\nu$ which in the interval
$-1\leq\nu<2$ is
given by the root of the function \cite{Petrov2}
\begin{equation}\label{lambda}
\lambda(\nu)=\frac{\nu(\nu+2)}{\nu+1}\cot\frac{\pi\nu}{2}+\frac{\nu\sin\gamma\cos(\nu\gamma+
\gamma)-\sin(\nu\gamma)}{(\nu+1)\sin^2\gamma\cos\gamma\sin(\pi\nu/2)}
\end{equation}
where $\gamma=\arcsin\left[M/(M+m)\right]$.
For equal masses we recover $s=2\nu+1\approx 2.55$ obtained in
Refs.~\cite{Petrov1,PSS}, and it slowly decreases with increasing the
mass ratio
(see Sec.~\ref{sec.relaxhl}).

The absolute value of the relaxation rate constant is determined by
the behavior
of the three-body system at short interatomic distances.
Assuming that the short-range physics is characterized by the
length scale $R_e$ and the energy scale $\hbar^2/mR_e^2$, we can
restore the dimensions and write:
\begin{equation}   \label{alphadd}
\alpha_{{\rm rel}}= C (\hbar R_e/m)(R_e/a)^s,
\end{equation}
where the coefficient $C$ depends on a particular system.

Experimental studies of bosonic  molecules produced in a Fermi gas
by using wide Feshbach resonances at JILA, Innsbruck, MIT, and ENS
are well described within the presented theoretical approach. The
JILA \cite{jila1,jila2,jila3}, Innsbruck \cite{rudy1,rudy2,rudy3},
MIT \cite{mit1,mit2}, ENS \cite{ens1,ens2}, and Rice \cite{randy2}
results show a remarkable collisional stability of weakly bound
molecules K$_2$ and Li$_2$ consisting of two fermionic atoms. At
molecular densities $n\sim 10^{13}$ cm$^{-3}$ the lifetime of the
gas ranges from tens of milliseconds to tens of seconds, depending
on the value of the scattering length $a$. A strong decrease of
the relaxation rate with increasing $a$, following from
Eq.~(\ref{alphadd}), is  consistent with experimental data. The
potassium experiment at JILA \cite{jila1} and lithium experiment
at ENS \cite{ens2} give the relaxation rate constant
$\alpha_{dd}\propto a^{-s}$, with $s\approx 2.3$  for K$_2$, and
$s\approx 1.9$ for Li$_2$, in agreement with our theory ($s\approx
2.55$) within
experimental uncertainty. The experimental and theoretical results
for lithium are shown in Fig.~4. The absolute value of the rate
constant for a $^6$Li$_2$ condensate is $\alpha_{dd}\approx
1\times 10^{-13}$ cm$^3$/s for the scattering length $a\approx
110\,$nm. For K$_2$ it is by an order of magnitude higher at the
same value of $a$ \cite{jila1}, which can be a consequence of a
larger value of the characteristic radius of interaction $R_e$.
\begin{figure}
\label{relax}
\includegraphics[width=6cm]{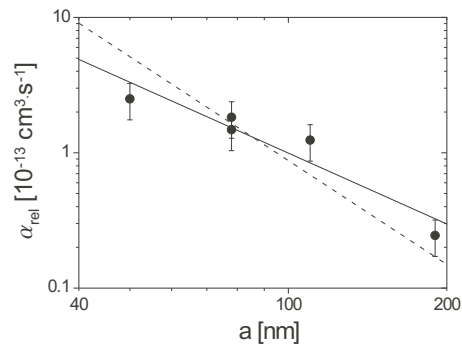}
\caption{Two-body decay rate $\alpha_{rel}$ of a  $^6$Li$_2$
molecular condensate a a function of interatomic scattering length
near the $^6$Li Feshbach resonance at 834\,G. Solid line: least
square fit, $\alpha_{rel}\propto a^{-1.9\pm 0.8}$. Dashed line,
theory: $\alpha_{rel}\propto a^{-2.55}$. The theoretical
relaxation rate has been normalized to the experimental value at
$a=78\,$nm.}
\end{figure}

\section{Collisional relaxation of molecules of heavy and light
fermions\label{sec.relaxhl}}

The analysis in the previous section also holds for collisional
relaxation of weakly
bound molecules of heavy and light fermionic atoms ($M/m\gg 1$). The
most efficient
is again the relaxation process in which one light and two heavy
fermions approach 
each other to distances $\sim R_e$, and the fourth (light) atom is
separated from
this three-body system by a large distance $\sim a$.

However, for a large mass ratio the behavior of the three-body system
of one light
and two heavy atoms changes, and the physical picture of the
relaxation process is
drastically modified. When the heavy atoms are separated from each
other by a
distance $x\ll a$, the light atom mediates an effective $1/x^2$
attraction between them.
This follows from the Efimov picture of effective interaction in a
three-body system
\cite{efimov} and can be illustrated in the Born-Oppenheimer
approximation, relying on the
results in Sec.~\ref{sec.bo} \cite{Fonseca}. For $x\ll a$, the light
atom is in
the bound state ``+'' with two heavy atoms, and from
Eqs.~(\ref{bo.lambda}), (\ref{epsilon}) 
we then obtain an effective potential $U_{eff}=\varepsilon_+\approx
-0.16\hbar^2/mx^2$. 
This attraction competes with the Pauli principle which in terms of
effective interaction 
manifests itself in the centrifugal $1/x^2$ repulsion between the
heavy atoms. 
The presence of this repulsion is clearly seen from the fact that in
the state ``+'' the
permutation of the heavy fermions does not change the sign of the
light-atom 
wavefunction $\psi_{{\bf x},+}({\bf y})$ given by
Eq.(\ref{bo.boundstatelight}).
Hence, the total wavefunction of the three-body system $\psi_{{\bf
x},+}({\bf y})\chi({\bf x})$
is antisymmetric with respect to this permutation only if the
wavefunction of the relative
motion of heavy atoms $\chi({\bf x})$ changes its sign. Therefore,
$\chi({\bf x})$ contains
partial waves with odd angular momenta, and for the lowest angular
momentum ($p$-wave) 
the centrifugal barrier is $U_c(x)=2\hbar^2/Mx^2$. For comparable
masses it is significantly 
stronger than $U_{eff}(x)$. Thus, we have the physical picture of the
previous section: the Pauli
principle (centrifugal barrier) reduces the probability for the
atoms to be at short distances and, as a consequence, the relaxation
rate decreases
with increasing the atom-atom scattering length $a$.

The role of the effective attraction increases with $M/m$. As a
result, the decrease
of the relaxation rate with increasing $a$ becomes weaker. In Fig.~5
we present the
dependence of the exponent $s$ in Eq.~(\ref{alphadd}) on the mass
ratio, obtained
from Eq.~(\ref{lambda}) which follows from the exact solution of the
three-body
problem. The exponent $s$ continuously decreases with increasing $M/m$
and becomes
zero for $M/m=12.33$. In the Born-Oppenheimer picture this means that
at this point
one has a balance between the mediated attraction and the centrifugal
repulsion.

\begin{figure}
\label{fig.nu}
\includegraphics[width=\hsize]{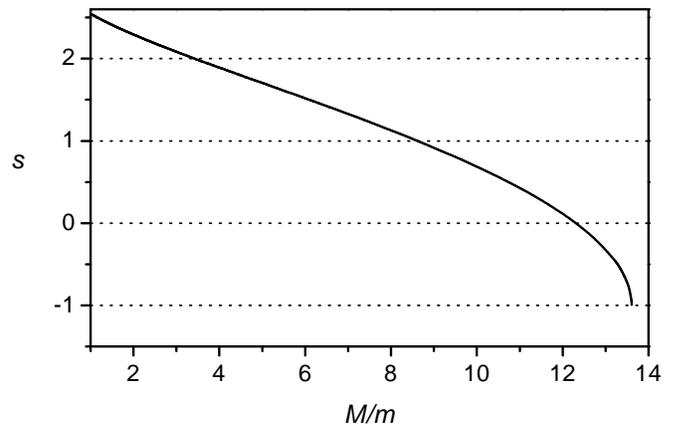}
\caption{The dependence of the exponent $s=2\nu+1$ in
Eq.~(\ref{alphadd}) on the mass
ratio $M/m$.}
\end{figure}

A further increase in $M/m$ makes $s$ negative and it reaches the
limiting value 
$s=-1$ for the critical mass ratio $M/m=13.6$. Thus, in the range
$12.33<M/m<13.6$
the relaxation rate increases with $a$.

For an overcritical mass ratio $M/m>13.6$ we have a well-known
phenomenon of the fall
of a particle to the center in an attractive $1/x^2$ potential
\cite{LL3}. As has
been mentioned in Sec.~\ref{sec.elastic}, in this case the shape of
the wavefunction 
at distances of the order of $R_e$ can significantly influence the
large-scale behavior and a short-range three-body parameter is
required to describe
the system.

The fact that for $M/m>12.33$ the relaxation rate is not decreasing
with an increase
in the atom-atom scattering length $a$, does not mean that the
relaxation is not
reduced compared to the case of molecules of bosonic atoms with the
same mass ratio.
For a large $M/m$, at distances $x$ larger than $a$ the molecules
interact via a
strong repulsive potential $U_{eff}(x)$ given by Eq.~(\ref{Ueff}).
Away from possible
resonances in molecule-molecule collisions, this barrier reduces the
amplitude of the
wavefunction at intermolecular distances $\sim a$ and, hence, leads to
a suppression
of the relaxation rate. Estimating the corresponding tunneling
probability $P$ in
the WKB approach, we obtain $P\propto \exp(-B\sqrt{M/m})$ where the
coefficient $B$
is of order unity. 

We thus see that one also expects the suppression of collisional
relaxation for
weakly bound molecules of light and heavy fermionic atoms. However,
the suppression
factor is independent of the atom-atom scattering length $a$ and is
governed by the
mass ratio $M/m$. The mechanism of this suppression originates from
Fermi statistics
for the light atoms, which leads to a strong repulsion between
molecules at large
intermolecular distances. This result should have implications for
experiments on
the formation of molecules in mixtures of $^6$Li with much heavier
fermionic atoms,
for example  $^{87}$Sr ($M/m\approx 14.5$) or $^{173}$Yb ($M/m\approx
29$).

\section{Molecular BEC and prospects for manipulations with weakly
bound molecules}

The suppression of the relaxation decay rate of weakly bound
molecules of fermionic atoms has a crucial con- sequence for the
physics of these molecules. At realistic temperatures the
relaxation rate constant $\alpha_{dd}$ is much smaller than the
rate constant of elastic collisions $8\pi a_{dd}^2v_T$, where
$v_T$ is the thermal velocity. For example, for the Li$_2$ weakly
bound molecules at a temperature $T\sim 3\mu$K and $a\sim 800$\AA,
the corresponding ratio is of the order of $10^{-4}$ or $10^{-5}$.
This opens wide possibilities for reaching BEC of the molecules
and cooling the Bose-condensed gas to temperatures of the order of
its chemical potential. Long-lived BEC of weakly bound molecules
has been recently observed for $^{40}$K$_{2}$ at JILA
\cite{jila2,jila3} and for $^{6}$Li$_{2}$ at Innsbruck
\cite{rudy2,rudy3}, MIT \cite{mit1,mit2}, ENS \cite{ens2}, and
Rice \cite{randy2}. Measurements  of the molecule-molecule
scattering length confirm the result $a_{dd}=0.6a$ with accuracy
up to 30\% \cite{rudy3,ens2}. This result is also confirmed by
recent Monte Carlo calculations \cite{giorgini} of the ground
state energy in the molecular BEC regime.

Remarkable achievements in the physics of ultracold Fermi gases
attract a lot of attention and draw fascinating prospects for
future studies. The prospects are to a large extent related to a
very long lifetime of weakly bound bosonic molecules of fermionic
atoms, which allows interesting manipulations with these
molecules. Arranging a deep evaporative cooling of their
Bose-condensed gas to temperatures of the order of the chemical
potential, one can then convert the molecular BEC into fermionic
atoms by adiabatically changing the scattering length to negative
values. This provides an additional cooling, and the obtained atomic
Fermi gas will have extremely low temperatures $T\sim 10^{-2}T_F$
which can be below the BCS transition temperature \cite{carr}. At
these temperatures one has a very strong Pauli blocking of elastic
collisions and expects the collisionless regime for the thermal
cloud, which is promising for identifying the BCS-paired state
through the observation of collective oscillations or free
expansion \cite{stringari,O'Hara,ens2,rudy3,Kinast}.

Recent experiments at Innsbruck, ENS, JILA and MIT have explored
the BEC-BCS crossover regime and shown a remarkably reversible
behavior of the gas from the BEC region to the BCS region and back
over duration of seconds \cite{rudy2,ens2,jila3,mit2, mit3}.
Experiments have also measured the pairing gap in the crossover
region using RF spectroscopy \cite{Chin04}. The energy gap varies
between the binding energy of the molecules in the region $a>0$, and
a density dependent (and much smaller) value in the region $a<0$
where no stable two-body bound state exists. Further experiments
at JILA and MIT have proven that fermionic  atom pairs present in
the strongly correlated regime at $a<0$  behave as Bose-condensed
pairs \cite{jila3,mit2,mit3}. For revealing the nature of this
superfluid pairing it is promising to make experiments with
rotating Fermi gases \cite{Sandro04} in the strongly interacting
regime, where one expects the formation of vortices and vortex
lattices \cite{Bulgac03,Feder04}. Another idea is to look directly
at a signature of the long range order in the superfluid phase by
an interference experiment as suggested in \cite{Carusotto04}.

It will also be interesting to transfer the weakly bound molecules
of fermionic atoms to their ground ro-vibrational state by using
two-photon spectroscopy, as proposed in Ref. \cite{DeM} for
molecules of bosonic atoms. Long lifetime of weakly bound
molecules of fermionic atoms at densities $\sim 10^{13}$ cm$^{-3}$
should provide a much more efficient production of ground state
molecules compared to the case of molecules of bosonic atoms,
where one has severe limitations on achievable densities and
lifetimes. One can then extensively study the physics of molecular
Bose-Einstein condensation. Moreover, heteronuclear molecules
which are supposed to be formed in mixtures of different fermionic
atoms, can be polarized by an electric field. One then gets a gas
of dipoles interacting via anisotropic long-range forces, which
drastically changes the physics of Bose-Einstein condensation
\cite{dipoles}.

\section*{Acknowledgments}

This work was supported by NSF through a grant for the Institute for
Theoretical
Atomic, Molecular and Optical Physics at Harvard University and
Smithsonian
Astrophysical Observatory, by the Minist\`ere de la Recherche (grant
ACI Nanoscience 201), by the Centre National de la Recherche
Scientifique (CNRS), and by the Nederlandse Stichting
voor Fundamenteel Onderzoek der Materie (FOM). LPTMS  is a mixed
research unit of CNRS and Universit\'e Paris Sud. LKB is a Unit\'e
de Recherche de l'Ecole Normale Sup\'erieure et de l'Universit\'e
Paris 6, associ\'ee au CNRS.

\end{document}